# Selective Octahedral Accommodation of $Cr^{3+}$ and Weak Magnetic Connectivity in the Sugilite Analogue $KNa_2Cr_2Li_3Si_{12}O_{30}$

**Yuya Haraguchi[1,*], Taishu Aoki[1], Daisuke Nishio-Hamane[2] and Hiroko Aruga Katori[1]**

[1]Graduate School of Engineering, Tokyo University of Agriculture and Technology, Koganei, Tokyo 184-8588

[2]The Institute for Solid State Physics, The University of Tokyo, Kashiwa, Chiba 277-8581, Japan

*Corresponding author: Email: chiyuya3@go.tuat.ac.jp

## Abstract

We report the synthesis of the Cr analogue of sugilite, $KNa_2Cr_2Li_3Si_{12}O_{30}$, in the milarite-type framework. Rietveld refinement of a composition-conserving antisite model gives $x$ = 0.0024(18) in $KNa_2[Cr_{2-x}Li_x][Li_{3-x}Cr_x]Si_{12}O_{30}$, corresponding to a $T2$-site Cr occupancy of 0.0008(6). X-ray MEM analysis shows no detectable Cr-like density at $T2$. Magnetic susceptibility indicates weak antiferromagnetic interactions with $\theta_W$ = −4.78(7) K and no ordering above 1.8 K.

### Graphical abstract

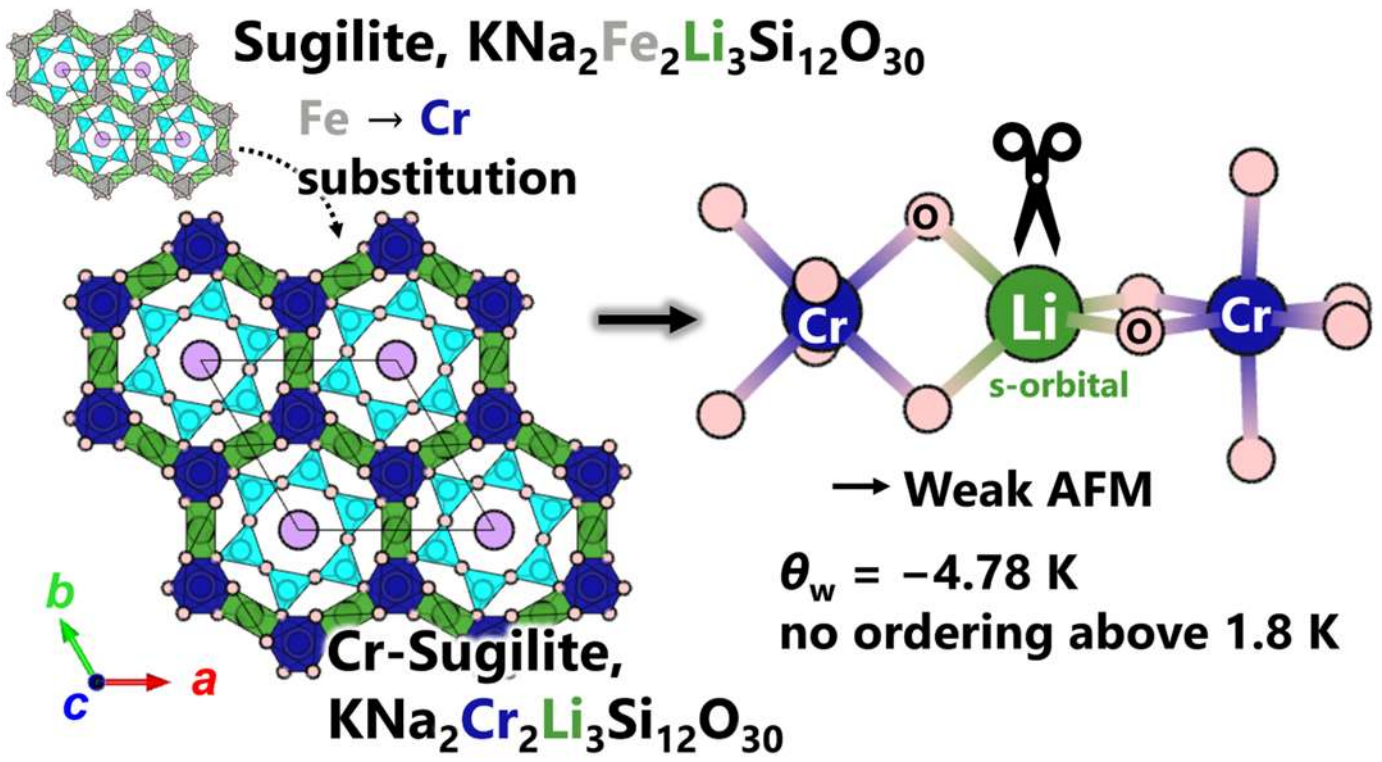

Mineral-based frameworks have already yielded some of the most influential model systems in quantum magnetism. A representative example is herbertsmithite, in which a mineral-derived kagome lattice gives one of the best-known candidates for a quantum spin liquid [1]. This success highlights the broader importance of mineral crystal chemistry as a guide for materials design, because naturally occurring framework topologies can provide structurally robust and geometrically distinctive magnetic sublattices [2-4].

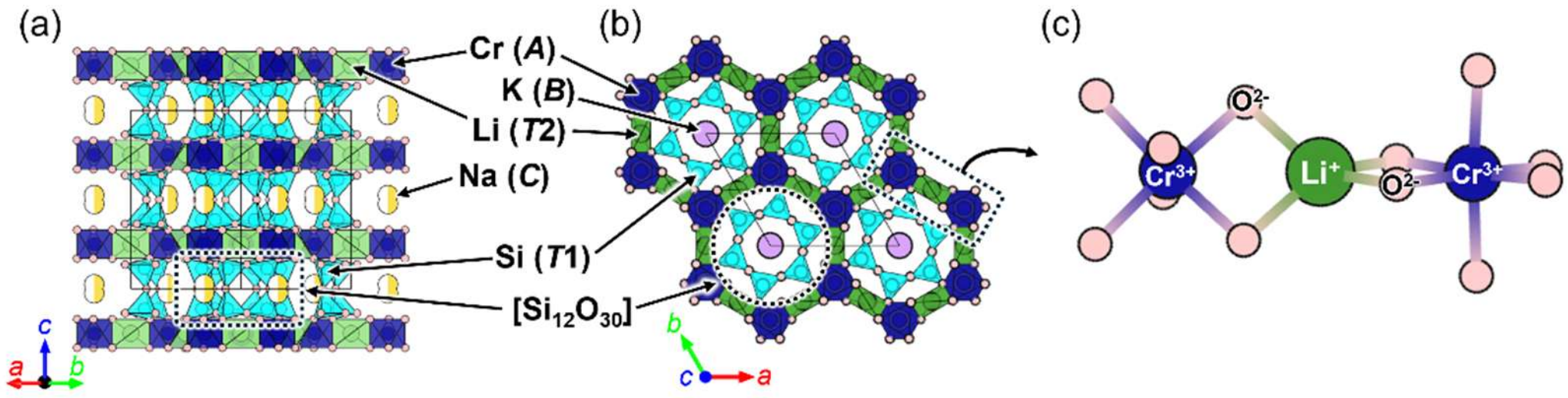

Fig. 1. Crystal structure of $KNa_2Cr_2Li_3Si_{12}O_{30}$. (a) Side view highlighting the layered framework separated by the $[Si_{12}O_{30}]$ double-ring unit. (b) Projection along the $c$ axis, showing the honeycomb-like arrangement of the $A$-site sublattice and the tetrahedral $T2$ sites linking neighboring $A$ sites. (c) Schematic illustration of a representative extended Cr-O-Li-O-Cr-type exchange path.

Table 1. Rietveld-refined atomic parameters for $KNa_2Cr_2Li_3Si_{12}O_{30}$ based on the antisite-disorder model $KNa_2[Cr_{2-x}Li_x][Li_{3-x}Cr_x]Si_{12}O_{30}$. The refined lattice parameters are $a = b = 10.003396(23)$ Å and $c = 14.037924(59)$ Å. The K, Na, Si, and O occupancies were fixed to the nominal crystallographic model, whereas the Cr/Li antisite parameter between the *A* and $T_2$ sites was refined. The antisite parameter converged to $x = 0.0024(18)$, corresponding to a T2-site Cr occupancy of 0.0008(6). The minor Cr component at $T_2$ shares the same 6f coordinates as the major Li component.

| label | site | *g* | *x* | *y* | *z* | *B* (Å$^2$) |
|---|---|---|---|---|---|---|
| K | 2*a* | 1 | 0 | 0 | 1/4 | 0.66(4) |
| Cr1 | 4*c* | 0.9988(9) | 1/3 | 2/3 | 1/4 | 0.92(2) |
| Li1 | 4*c* | 0.0012(9) | 1/3 | 2/3 | 1/4 | 0.92(2) |
| Li2 | 6*f* | 0.9992(6) | 0.5 | 1/2 | 1/4 | 0.5(fix) |
| Cr2 | 6*f* | 0.0008(6) | 0.5 | 1/2 | 1/4 | 0.5(fix) |
| Si | 24*m* | 1 | 0.23687(6) | 0.35855(6) | 0.38716(4) | 0.52(9) |
| O1 | 12*l* | 1 | 0.14136(17) | 0.39922(17) | 0 | 0.59(4) |
| O2 | 24*m* | 1 | 0.22326(13) | 0.27557(11) | 0.13623(7) | 0.86(3) |
| O3 | 24*m* | 1 | 0.17028(12) | 0.51156(13) | 0.17011(6) | 0.94(3) |
| Na | 8*h* | 0.5 | 1/3 | 2/3 | 0.02528(14) | 1.08(9) |

Mineral-derived double-ring silicates are attractive in this context because they contain crystallographically distinct cation sites within a rigid silicate scaffold. In the milarite-group, represented by the general formula $(C)(B)_2(A)_2(T2)_3(T1)_{12}O_{30}$, the A site is octahedral and the T2 site is tetrahedral, whereas the *T*1 site is mainly occupied by Si in the $[Si_{12}O_{30}]$ double-ring framework [5-8]. As illustrated in Figure 1(a), the structure consists of polyhedral layers separated along the c axis by rigid $[Si_{12}O_{30}]$ double-ring silicate units, with K occupying the *C* site at the centers of the rings and Na occupying the *B* site between adjacent rings. When projected along the c axis [Figure 1(b)], the octahedral *A*-site sublattice forms a honeycomb-like arrangement, whereas the tetrahedral *T*2 sites occupy the links between neighboring *A* sites. This means that the framework combines a clear distinction between octahedral and tetrahedral coordination environments with a structurally layered architecture. At the same time, the projected honeycomb-like topology does not imply a short direct exchange path, because neighboring *A* sites are connected only through intervening tetrahedra and oxygen bridges, as shown schematically in Figure 1(c). The milarite-group is therefore a suitable platform for testing whether crystal-chemically driven site selectivity can be used to define a magnetic sublattice in a controlled manner.

The present target compound is the Cr analogue of sugilite, $KNa_2Cr_2Li_3Si_{12}O_{30}$. Conventional sugilite is $KNa_2Fe_2Li_3Si_{12}O_{30}$ [5,8]. Replacing $Fe^{3+}$ with $Cr^{3+}$ is appealing because $Cr^{3+}$ is expected to prefer octahedral coordination much more strongly than tetrahedral coordination on crystal-chemical grounds [6,9]. If this preference dominates in the sugilite- type framework, Cr should partition almost entirely into the octahedral A site, forcing Li into the tetrahedral *T*2 site and sharply suppressing *A*/*T*2 antisite exchange. The significance of this substitution is therefore not merely compositional. Rather, Fe-to-Cr substitution offers the possibility of converting an inferred site assignment into an experimentally explicit structural feature.

Polycrystalline $KNa_2Cr_2Li_3Si_{12}O_{30}$ was synthesized by a conventional solid-state reaction from $K_2CO_3$ (99.9%), $Na_2CO_3$ (99.5%), $Cr_2O_3$ (99.9%), $Li_2CO_3$ (99.9%), and SiO2 (99.99%) (all from Kojundo Chemical Lab. Co., Ltd.). The mixed powders were pelletized and heated in air at 950°C for 24 h, followed by two regrinding and reheating cycles at 1000°C for 48 h each. Powder X-ray diffraction data were collected with Cu *K*α radiation on a Rigaku MiniFlex600 diffractometer over 5-100° in $2\theta$ with a step width of 0.01°. Rietveld refinement was carried out using Z-Rietveld [10]. To examine the electron-density distribution, maximum entropy method analysis was performed from the refined diffraction data [11]. Magnetic susceptibility was measured with an MPMS magnetometer under 0.1 T in the temperature range 1.8-300 K. Chemical analyses were conducted using a scanning electron microscope equipped with an energy-dispersive X-ray spectrometer (SEM-EDX; JEOL IT-100) on selected Cr-containing main-phase grains to evaluate the K, Na, Cr, and Si contents and compositional homogeneity. Li was not quantified by EDX because of its intrinsically weak X-ray signal.

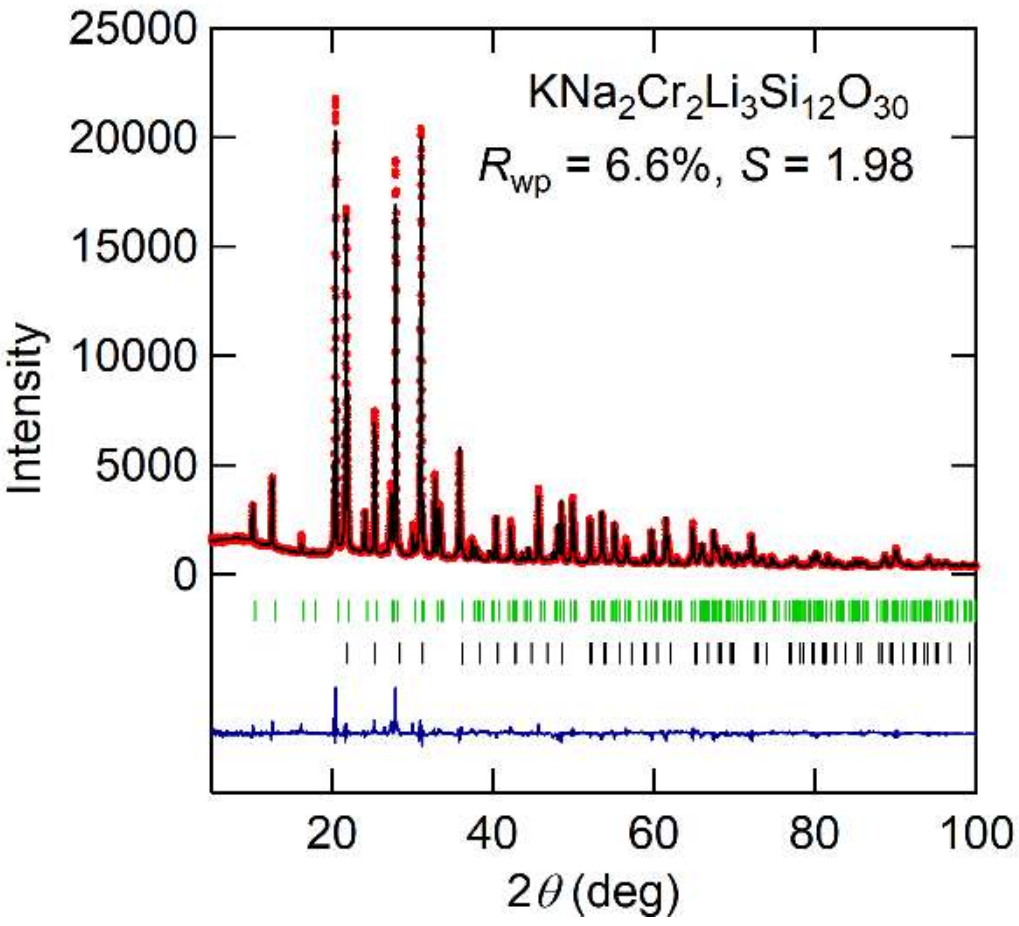

Fig. 2. Powder X-ray diffraction pattern and Rietveld refinement profile of $KNa_2Cr_2Li_3Si_{12}O_{30}$. The observed, calculated, and difference patterns are shown together with the Bragg positions of the main phase and the minor α-cristobalite impurity.

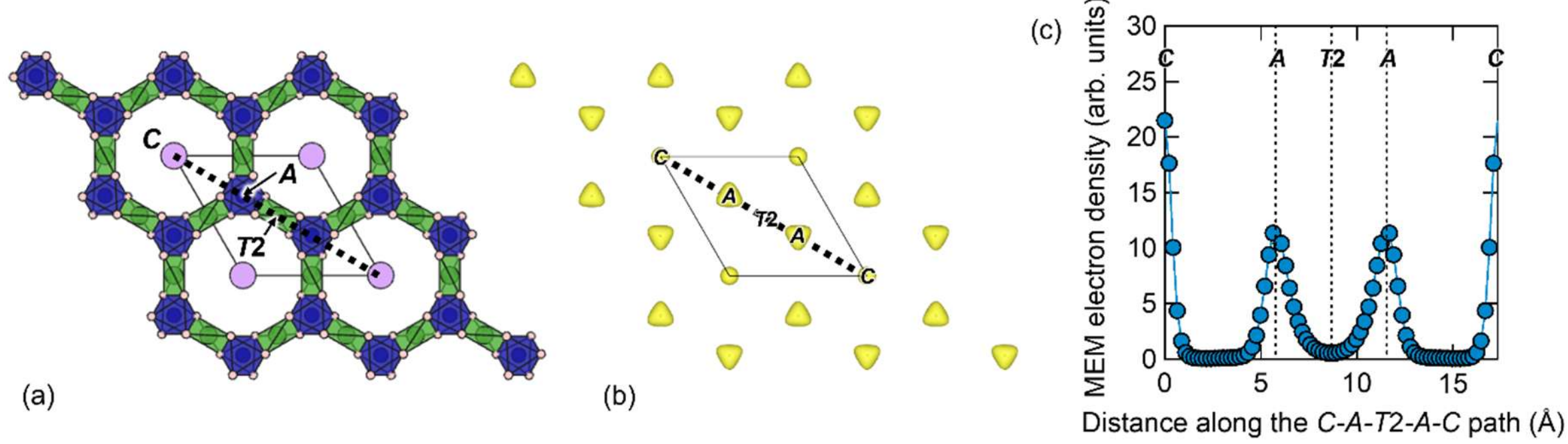

**Fig. 3**. Maximum entropy analysis (MEM) of $KNa_2Cr_2Li_3Si_{12}O_{30}$. (a) Structural projection indicating the *C*, *A*, and *T2* sites. (b) Corresponding MEM electron-density map. (c) Line profile of the MEM electron density along the *C-A-T2-A-C* path, showing strong density maxima near *A* and no comparable central peak at *T2*.

The powder x-ray diffraction pattern is shown in Fig. 2. The major reflections are indexed by the sugilite-type structure with space group *P*6/*mcc*. The fit improved substantially when alpha-cristobalite was included as a minor impurity phase. The final two-phase refinement gave $R_{wp}$ = 6.6% and $S$ = 1.98, with lattice parameters $a = b$ = 10.003 396(23) Å and $c$ = 14.037 924(59) Å. The $SiO_2$ impurity fraction was estimated as 3.97 wt%. The refined $c$ axis corresponds to $c/2$ = 7.02 Å for adjacent structural layers, consistent with substantial structural separation imposed by the $[Si_{12}O_{30}]$ double-ring silicate unit. The presence of α-cristobalite is also chemically reasonable because cristobalite formation can be promoted in alkali-containing $SiO_2$ systems at relatively low temperatures [12]. To evaluate possible alkali loss during the high-temperature reaction, SEM-EDX point analyses were performed on Cr-containing main-phase grains. Because Li cannot be quantified by EDX, the analysis was used to examine the EDX-detectable cations. Normalization to Si = 12 gave average values of Na = 1.97(11), K = 0.77(6), and Cr = 2.01(5). Thus, Na and Cr are close to the nominal values, and no obvious Na deficiency or Cr/Si compositional anomaly was detected in the main phase. The K content is lower than the nominal value in the SEM-EDX analysis. Since EDX does not quantify Li and chemical composition does not directly determine crystallographic site occupancies, the EDX result was not used to refine absolute alkali-site occupancies. Instead, it was used as a composition check for the detectable cations. The individual analyses and representative BSE images are provided in Table S1 and Figure S1 in the Supplementary Material.

The central structural question is whether Cr is selectively accommodated at the octahedral *A* site or whether appreciable antisite occupation occurs at the tetrahedral *T2* site. To test this directly, the diffraction data were refined using a composition-conserving model, $KNa_2[Cr_{2-x}Li_x][Li_{3-x}Cr_x]Si_{12}O_{30}$, where x represents Cr/Li exchange between the *A* and *T2* sites. The refinement yielded $x$ = 0.0024(18), corresponding to a *T2*-site Cr occupancy of only 0.0008(6). Thus, antisite exchange between the *A* and *T2* sites is negligible within experimental uncertainty, demonstrating essentially complete partitioning of Cr into the octahedral *A* site. Powder X-ray diffraction and EDX do not independently determine the absolute Li occupancy. Nevertheless, because the refined Cr occupancy at $T_2$ is only 0.0008(6), the $T_2$ site is assigned predominantly to Li on the basis of the nominal synthesis composition, crystallographic site multiplicity, and the expected site preference of $Cr^{3+}$.

This conclusion is independently supported by MEM analysis. Because X-ray MEM is relatively insensitive to Li, the purpose of the MEM analysis was not to identify Li directly, but to test whether any appreciable Cr-like density is present at *T2*. Figures 3(a) and 3(b) show the projected structure and the corresponding MEM map, and Figure 3(c) gives the MEM line profile along the *C-A-T2-A-C* path. Pronounced density maxima are observed in the vicinity of the A sites, whereas no comparable central peak attributable to Cr is found at the *T2*-site center. This result is fully consistent with the refined *T2*-site Cr occupancy of 0.0008(6). The present Cr analogue therefore makes the *A*/*T2* partitioning experimentally explicit within the sugilite-type framework.

The significance of the present compound becomes clearer when it is compared with conventional sugilite. In modern structural treatments of sugilite, Fe-rich trivalent cations are assigned to the octahedral *A* site and Li to the tetrahedral *T2* site, and recent work has focused mainly on alkali-ion behavior rather than substantial Fe/*T*2 antisite exchange [7,8]. In the present Cr analogue, by contrast, the *A*/*T*2 partitioning is not merely inferred from crystal chemistry but is directly supported by both unconstrained occupancy refinement and MEM analysis. In this sense, Fe-to-Cr substitution does not change the underlying sugilite topology but turns a crystal-chemical site assignment into a directly testable structural feature.

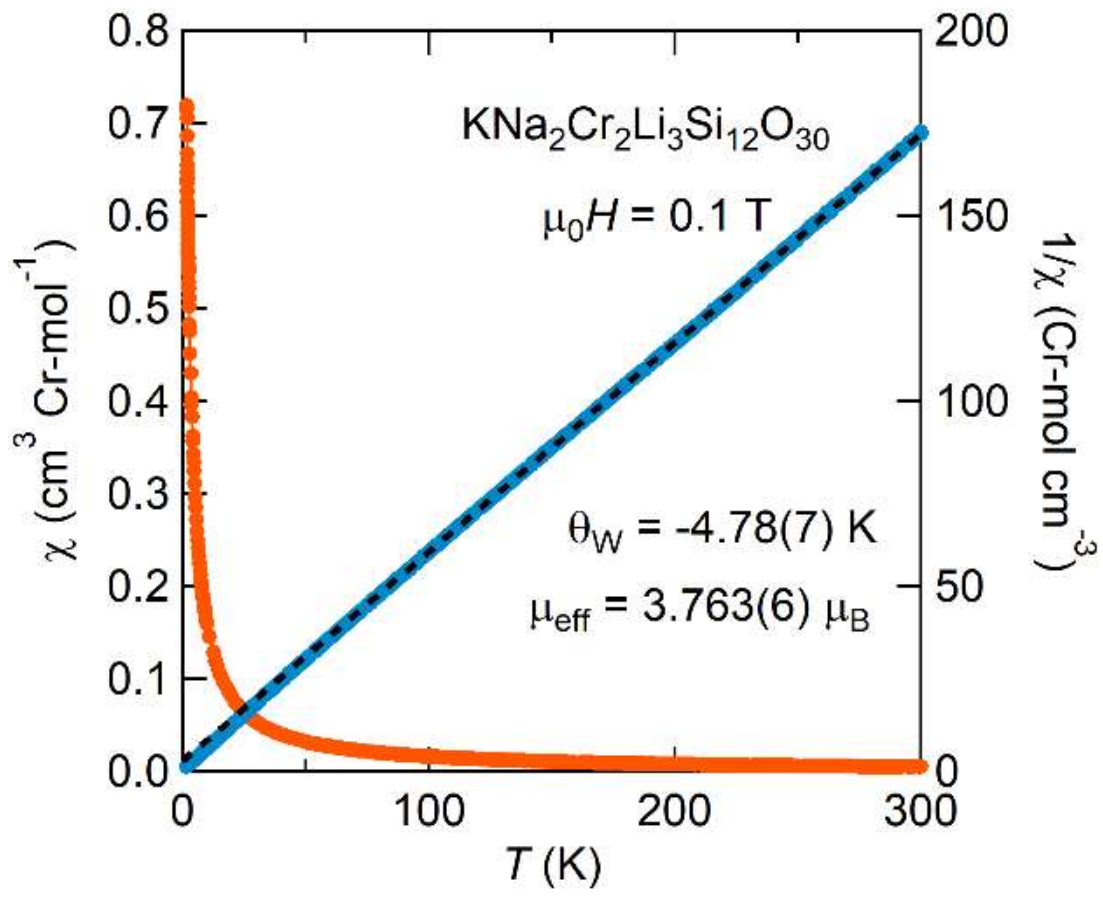

**Fig. 4**. Temperature dependence of the magnetic susceptibility and inverse susceptibility of $KNa_2Cr_2Li_3Si_{12}O_{30}$ measured at 0.1 T. The broken line indicates the Curie-Weiss fit over 100-300 K.

The refined structure implies that the Cr sublattice corresponds to the *A*-site honeycomb-like arrangement in projection, as shown in Figure 1(b). However, this should not be taken as evidence for a strongly interacting honeycomb magnet. In the present structure, no direct Cr-O-Cr superexchange pathway is available between neighboring *A*-site Cr centers. Instead, magnetic coupling must be transmitted through an extended Cr-O-Li-O-Cr linkage as shown in Figure 1(c). Because $Li^+$ is a closed-shell ion and does not provide an efficient virtual-hopping channel, superexchange along this pathway is expected to be strongly suppressed. In this sense, the projected honeycomb-like Cr network is structurally defined but magnetically only weakly connected. This interpretation is consistent with the magnetic susceptibility shown in Figure 4. The Curie-Weiss fit over 100–300K gives $\theta_W$ = −4.78(7) K and $\mu_{eff}$ = 3.763(6) $\mu_B$ per Cr, with no sign of magnetic ordering above 1.8 K. Additional fits over 150–300 K and 200–300 K give similarly small $\theta_W$ values of −4.22(12) K and −2.67(20) K, respectively, indicating that the conclusion of weak magnetic coupling does not depend strongly on the fitting range. The observed $\mu_{eff}$ is also close to the spin-only value expected for $Cr^{3+}$ with S = 3/2. In the refined average structure, the $CrO_6$ octahedron is nearly regular, with six Cr-O distances of approximately 1.948 Å. Thus, no large distortion of the octahedral crystal field is indicated. For octahedrally coordinated $Cr^{3+}$, the $3d^3$ configuration gives a $t_{2g}^3$, S = 3/2 state with a largely quenched orbital contribution. and the observed magnetic moment is consistent with the trivalent Cr assignment at the A site.

$KNa_2Cr_2Li_3Si_{12}O_{30}$ is therefore best viewed as a structurally clean realization of explicit *A*/*T*2 site partitioning in a milarite-type framework. The main outcome is not the discovery of an unusually strong honeycomb or kagome magnet, but the demonstration that the *A*/*T*2 partitioning in this structure can be sharpened by appropriate cation choice and verified directly by occupancy refinement and maximum entropy analysis. At the same time, the present result provides design rules for future milarite-type magnets along two complementary directions. For honeycomb magnets based on the A-site sublattice, the present compound shows that selecting a magnetic ion with strong octahedral preference is not sufficient by itself: the *T*2 tetrahedron must also be redesigned, because Li-dominant *T*2 chemistry is too electronically inert to support substantial *A*-O-*T*2-O-*A* super-superexchange. Conversely, kagome magnets should be pursued by fixing the octahedral *A* site with diamagnetic cations and activating the *T*2 kagome-like sublattice by introducing tetrahedrally compatible magnetic ions there. In this sense, the present Cr analogue is not only a reference system for explicit site partitioning, but also a practical guide for designing both honeycomb and kagome magnets in the milarite-type framework.

**Acknowledgement**

Part of this work was carried out by joint research in the Institute for Solid State Physics, the University of Tokyo (Project Numbers 202311-MCBXG-0021, 202311-MCBXG-0025, 202406-MCBXG-0100, 202406-GNBXX-0095, 202406-MCBXG-0100, 202406-MCBXG-0101, 202411-MCBXG-0033, and 202411-MCBXG-0034).

**Funding**

This work was supported by JST PRESTO Grant Number JPMJPR23Q8 (Creation of Future Materials by Expanding Materials Exploration Space) and JSPS KAKENHI Grant Numbers. JP25K01496 (Scientific Research (B)), JP23H04616 and JP25H01649 (Transformative Research Areas (A) "Supra-ceramics"), JP25H01403 (Transformative Research Areas (B) "Multiply Programmed Layers"), JP22K14002 (Young Scientific Research), and JP24K06953 (Scientific Research (C)).

*Conflict of interest statement.* None declared.